\documentclass[prb,aps,twocolumn]{revtex4}
\usepackage{epsfig}
\usepackage{longtable}
\begin{document}
\title{ Molecular dynamics study of the   
fragmentation of silicon doped fullerenes}
\author{Chu-Chun Fu\cite{fr}, Javier Fava, Ruben Weht, M.Weissmann}
\affiliation{Departamento de F\'{\i}sica, Comisi\'on Nacional de Energ\'{\i}a 
At\'omica, Avda. del Libertador 8250, 1429 Buenos Aires, Argentina}

\begin{abstract}
Tight binding molecular dynamics simulations, with a non orthogonal basis
set, are performed to study the fragmentation of carbon fullerenes doped 
with up to six silicon atoms.
Both substitutional and adsorbed cases are considered.
The fragmentation process is simulated starting from the equilibrium
configuration in each case and imposing a high initial 
temperature to the atoms.  Kinetic energy quickly converts
into potential energy, so that the system  oscillates for some picoseconds
and eventually  breaks up.
The most probable first event for substituted fullerenes is the ejection
of a $C_2$ molecule,  another very frequent event being that one $Si$
atom goes to an adsorbed position.
Adsorbed $Si$ clusters  tend to desorb as a whole when they have four or more
atoms, while the smaller ones tend to dissociate and sometimes interchange
positions with the $C$ atoms.
These results are compared with experimental information from mass abundance
spectroscopy and the products of photofragmentation.
\end{abstract}

\maketitle

\section{Introduction}

Since the discovery of $C_{60}$ in 1985 \cite{kroto} 
and the following large scale
synthesis of fullerenes, considerable effort has been devoted to use these
molecules as building blocks for novel and more interesting materials. One
of the most important attempts is to try to change their electronic and
mechanical properties through
doping. In recent years exohedrally and endohedrally doped fullerenes have
been produced, with a variety of doping atoms and even small molecules.
 In the exohedral case, the
foreign atoms (or groups of atoms) are outside, attached to the fullerene, as
in the case of the superconducting fullerides. In the endohedral case the
molecule encloses the strange element, isolating it from the outside.
Many possible applications, specially in medicine, have been envisaged using
these new systems. A third approach to create fullerene-related compounds
is to modify the fullerene itself through substitutional doping. This is
expected to be easier if the dopant atom has a similar electronic
configuration to that of carbon, but has been successfully performed with
several elements, such as B, N, O and some transition metal atoms as Fe, Co,
Ni, Rh and  Ir.
 At first glance, substitutional doping with silicon should be
also easy due to the similar electronic configuration of carbon and silicon.
However, both atoms prefer quite different kinds of bonding. While carbon can
form $sp^1, sp^2$, and $sp^3$  bonds, silicon strongly prefers the
$sp^3$ configuration, making silicon clusters to be generally 
in a three-dimensional
arrangement. Silicon doped and silicon coated fullerenes have been
produced only very recently
\cite{kimura,fye,pellarin1,pellarin2,pellarin3}, starting either from an initially doped
carbon-based material or from precursor pure carbon fullerenes interacting 
with a vapor of the doping element. Both mass abundance spectroscopy and the
analysis of the photofragmentation products of selected species show clearly
that several carbon atoms can be substituted in the cage structure and also
that silicon clusters can be adsorbed on its surface.

From the theoretical side, some semiempirical \cite{menon} and
ab-initio calculations \cite{parrinello,lu,japoneses,nuestro} 
have been performed to study the
structure, stability  and electronic properties
of these new molecules.  Up to now the main way 
 to infer the structure of the doped systems is through the study
of their fragmentation and the analysis of the residual clusters with
different compositions. For this reason, 
in the present work we attempt to model those fragmentation processes,
using molecular dynamics simulations within a density functional based 
tight-binding model \cite{porezag,frauenheim}.
This seems to be an adequate choice, as it takes into account the bonding
characteristics of the two elements, carbon and silicon, being less
time consuming than full ab-initio simulations.

This paper is organized as follows: In  section II we present the
method used in the calculations, putting special emphasis on the different
approximations used. In section III we analize the results, first for
substitutional doping and then for coated fullerenes. Finally in section
 IV are our conclusions and comments.

\section{Method of calculation}

Along this work we use a density functional based
non orthogonal tight binding hamiltonian, that was developed and tried
for different systems containing carbon and silicon atoms 
\cite{porezag,frauenheim}.

In this method the hamiltonian and overlap matrix elements were obtained
from pseudo-atomic orbitals as a function of distance, in a parameter free way.
In this sense it seems more satisfactory than the orthogonal 
tight-binding parametrization we previously developed for mixed systems 
\cite{nuestroprevio}.  Although only a minimal basis set is used,
the method has been proved to be transferable, giving good results
for clusters, surfaces and solids \cite{frauenheim}.  
It gives also a good description of some non-trivial low dimensional systems, 
as for example the reconstructed Si and $\beta-$SiC surfaces \cite{eduardo}, 
which is an indication that a reasonable description of the 
fullerenes when bonds break up should be obtained.  

The method uses only  two center integrals and the total  
energy is written as the sum of a band-structure term and a repulsive term,
this last one being parametrized also with experimental information.
When performing the dynamical simulations at finite temperatures or obtaining 
the relaxed equilibrium structures, the atoms are moved according to 
the Hellmann-Feynmann forces
and the equations of motion are integrated using Verlet's algorithm.

The approach used here does not consider charge self-consistency to avoid
unreasonably large simulation times and the introduction of a new set 
of parameters (a Hubbard type U and a Madelung correction).
The effect of this approximation is that charge transfers may be rather
large and the fragments carry incorrect amount of net charges. 
Semiempirical methods, like this one, should be used instead of an ab-initio 
calculation when this one makes the simulation too lengthy.

It is worth mentioning that experiments are performed on single positively
charged clusters, however the fragmentation of non-doped
fullerenes, the succesive emission of $C_2$ molecules is
correctly described by  simulations with neutral molecules. For this
reason  we have used
the same approximation in this case. Nevertheless, we have also proved with a 
few examples that
there is no statistical difference when using positive ions instead of
neutral fullerenes if the same parameters are used.
Our previous ab-initio calculations \cite{nuestroprevio} also show that even  
in the case of Si substituted ions, where the original fullerene cage is 
far more strongly perturbed than in the adsorbed case, we do not find 
very different geometries or energy gaps. 
We therefore do not expect the fragmentation results to differ in any 
important way between the neutral molecules and the positive ions, in both
substitutional and coated cases.

Before starting  the simulations we  checked that the structure,
stability and electronic properties of the substituted fullerenes, that we
studied previously with an ab-initio procedure \cite{nuestro}, are correctly
reproduced.  In fact, the substituted fullerenes
are also stable in this case and the energy ordering of the
different isomers containing 2, 3 and 6 carbon atoms substituted by 
silicon atoms is well reproduced, giving energy differences  of the same
order of magnitude.    The energy differences obtained by the
tight-binding
method are not consistently higher or lower than the corresponding ab-initio
ones. The two characteristic $C-C$ bond lengths in the fullerene cage are
maintained in the relaxed substituted molecules and the lowest energy, when
more than two silicon atoms are in the cage, is when they are nearest
neighbors, so as to reduce the number of $Si-C$ bonds. As in the ab-initio
calculations, sometimes lower energy configurations present some weak bonds that
could indicate the possible fragmentation paths.
In the case of adsorbed
silicon atoms,  they form weakly
bound clusters of silicon atoms on the surface of the fullerene molecule, with
silicon atoms locating preferentialy in front of a hh (hexagon-hexagon)
 $C-C$ bonds
and close to each other for cases with more than two adsorbed silicon atoms. 

With respect to the electronic properties the tight-binding calculations
give very similar values for the band-gaps but,
as expected for a non self-consistent calculation, the charge transfers
between different atoms are rather
large, about double those of the ab-initio calculations, and the bond orders
somewhat smaller.

To simulate the fragmentation process we start from a relaxed molecule
at 0 K and attribute a random velocity to each atom, with a maxwellian
distribution corresponding to a high  temperature. 
It assumes a very fast energy 
interchange between electronic and ionic degrees of freedom.
This high initial ionic temperature approximation
 has been used previously to simulate the  excitation produced
by a femtosecond laser pulse \cite{scuseria}. 
In these experiments, after 
being photoexcited
the molecules do not collide before dissociation, so that it is quite
reasonable to assume energy conservation during the fragmentation process.
For pure fullerenes a kinetic energy corresponding to an initial
temperature of about 12000 K \cite{scuseria}  was needed
to produce the first breakup, that is, the first  emission of a $C_2$
molecule, in a reasonable simulation time of tens of picoseconds. 
This time is of course much
smaller than the experimental time of flight, but the experimental facts
are correctly reproduced and the energies involved  in reasonable agreement 
\cite{tomita}. The occupation of the electronic
energy levels is time dependent, in each simulation step it corresponds to
the Fermi function of the corresponding instantaneous ionic temperature. However,
due to the  large energy gap of the fullerene molecule, this
only allows for a  small amount of electronic excitation to the antibonding
states. 
The microcanonical molecular dynamics interchanges kinetic and
potential energy so that after a few time steps the system equilibrates at
about 3000 K in the silicon substituted cases and it vibrates for some
picoseconds before breaking up,
the initial energy converts quickly into molecular vibrations.

 Substituted fullerenes need lower initial temperatures to break up, and
coated ones even lower, but the contribution of excited states
is always quite small, especially once the systems achieve a stable 
temperature of 3000K or less. 
According to our previous ab-initio calculations and also to the present 
tight binding ones the Si doped fullerenes do not have a much smaller
energy gap than the unsubstituted molecule. 
For example it is 1.32 $eV$ 
for $C_{59}Si$ and around 1 $eV$ for the different isomers of $C_{57}Si_3$, 

Each simulation run in this work lasted between
20 and 30 ps,   the time step
being either 0.5 fs or 1fs. This assures a reasonable energy conservation 
for the whole simulation. Due to the random nature of the initial conditions,
several examples were performed for each type of molecule and each initial
temperature.

\section{Results}
\subsection{Substituted Fullerenes}

The tight binding approximation is only valid when the leaving fragment
is at a smaller distance than the cut off from the remaining molecule.
We therefore report here only the first
event observed in each simulation run, and attribute significance to 
the results up to that fragmentation instant.  Once the 
system separates into fragments, they do not interact with each other and 
therefore the tight-binding wavefunction does not make sense.

After the high initial temperature is applied, between 6000 and 9000 K,
the  system achieves in less than one picosecond a temperature of 
around 3000 K 
due to the fast interconversion between kinetic and
potential energy.

\begin{figure}[b]
\epsfxsize=6.6cm\centerline{\epsffile{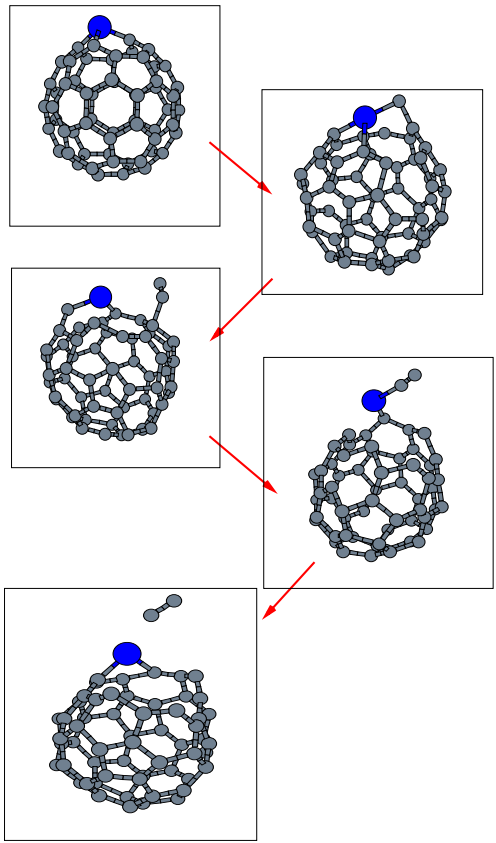}}
\caption{Selected instantaneous intermediate structures, leading to the
fragmentation of $C_{59}Si$}
\label{Fig.1}
\end{figure}

When more than one silicon atom is present, the results may differ for
the different isomers. In some cases it is particularly easy for one
silicon atom to become adsorbed outside of the carbon cage, in others
it is not. The dynamics of three different isomers is 
 studied for two substituted
atoms, two different isomers in the case of three substitutions and
only the lowest energy isomer is studied in the case of six substitutions.
Table 1 shows the results from
 all the  simulations, totalling about 3300 ps, and the following
general features can be observed:

1) The most frequent first event is the ejection of a $C_2$ molecule, usually
located near a silicon atom. The first bond to break is one between a
 $Si$ and a $C$ atom, which leaved an open chain of carbon atoms from which
$C_2$ is removed. The $Si$ atom  rebonds quickly, as it strongly prefers
tetrahedral coordination, while the carbon atoms are quite confortable in the
chain structure $sp^1$ (see Fig.1 ).

2) Another very frequent first event is  that a $Si$ substituted atom moves
to an adsorbed position, from which it  orbits around the
fullerene cage. This adsorbed atom easily leaves the molecule, alone.
(see Fig. 2)

\begin{figure}[b]
\epsfxsize=6.6cm\centerline{\epsffile{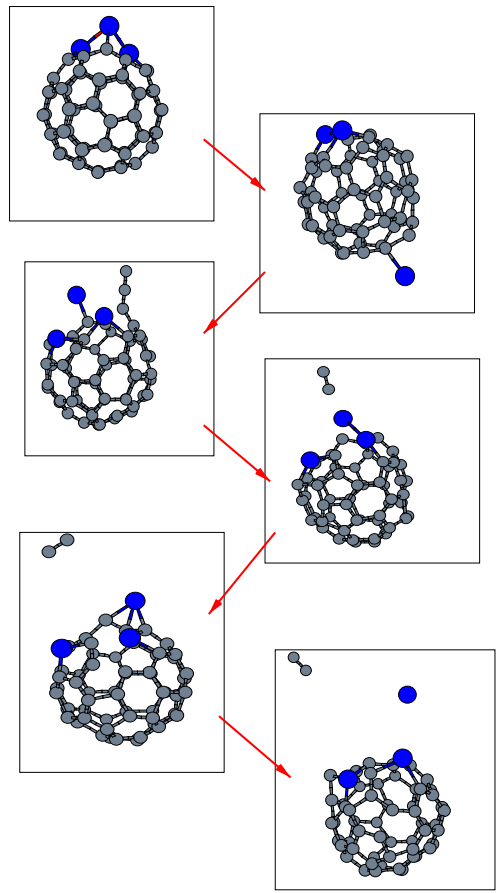}}
\caption{Same as Fig.1, for one of the observed paths leading to the
fragmentation of $C_{59}Si_3$}
\label{Fig.2}
\end{figure}

\begin{table}[!]
\caption{First events observed in the fragmentation of $Si$ substituted 
fullerenes. The maximum 
simulation time is 30 ps for each run and between 10 and 20 simulations
were ran for each system. 
The more stable isomers are indicated with an asterisk, and $nn$ indicates
that the $Si$ atoms are nearest neighbors.\\}  
\begin{tabular}{cccc}
System & Initial & Freq. & Event description \\
       & Temp.   &       &                   \\ \hline \hline
& & & \\
$C_{59}Si$     &9000 K& & \cr
 &&0.82 & $C_2$ ejected  \cr
 &&0.06 & $SiC_2$ ejected \cr
 &&0.12 & no frag. observed \cr
 & &&  \cr
$C_{59}Si$ &8000 K&  &   \cr
 &&0.50& $C_2$ ejected  \cr
 &&0.08 &$Si$ ejected  \cr
 &&0.08 &$C_4$ ejected  \cr
 &&0.34 & no frag. observed  \cr
 && &  \cr
$C_{58}Si_2$ & 8000 K& &   \cr
$Si$ in a pentagon && 0.10& $C_2$ ejected   \cr
($nn$) && 0.10& $Si$ ejected \cr
 && 0.60& $Si$ became adsorbed \cr
 && 0.20& no frag. observed \cr
 && &  \cr
$C_{58}Si_2$ & 8000 K&  &  \cr
$Si$ in a pentagon &&0.90& $C_2$ ejected \cr
 (not $nn$) &&0.05 & $SiC_2$ ejected  \cr
 &&0.05 & no frag. observed \cr
 && &  \cr
$C_{58}Si_2$ & 8000 K&  &   \cr
$Si$ in a hexagon && 0.67 & $C_2$ ejected  \cr
 (opposite) $*$ && 0.33 & no frag. observed \cr
 && &  \cr
$C_{57}Si_3$ & 6500 K& &   \cr
$Si$ in a hexagon && 0.30& $C_2$ ejected  \cr
($nn$) $*$ && 0.50& adsorbed $Si$ ejected \cr
 && 0.20&  no frag. observed \cr
 && & \cr
$C_{57}Si_3$ &6500 K&  &   \cr
$Si$ in a pentagon& & 0.20& $C_2$ ejected \cr
 (not $nn$) $*$& & 0.10& $SiC_2$ ejected \cr
 && 0.10& $Si$ became adsorbed \cr
 && 0.60& no frag. observed \cr
 & && \cr
$C_{54}Si_6$ & 6000 K&  &  \cr
$Si$  in a hexagon $*$& & 0.50& $C_2$ ejected \cr
& & 0.10& $Si_2C$ ejected\cr
& & 0.10 & $Si$ ejected \cr
& & 0.05 & $Si_2$ ejected \cr
& & 0.25 & no frag. observed \\ \hline \hline
\end{tabular}
\end{table}

3) Odd numbered molecules, for example those that remain after one $Si$ atom 
leaves, seem to be
as stable as even numbered ones within our simulation time, although
the calculated binding energy is smaller for the odd molecules.
However, the surviving
ionic products in the mass spectra 
show very few molecules with an odd number of atoms.

4) Small clusters containing both $Si$ and $C$ atoms are ejected in a few cases
but again there is no significative preference for even numbers.

The lack of self-consistency may be one  reason for not finding the
molecule $SiC$ as a result of fragmentation, as it appears in the experimental 
results.
Other possible explanations for the above mentioned discrepancy between
experimental and simulation results can be suggested. For example, the
experimental laser excitation lasts nanoseconds, therefore allowing for
sequential absorbtion of several photons,  while the simulations only
study fast femtosecond laser induced fragmentations. 
 Different time scale excitations could  produce different fragments.

Also, the simulations study essentialy the dynamics of the first fragmentation
event but the
experimental spectra may be more related to the energies of the final products.
For example, starting from $C_{59}Si$ one may end up with  $C_{58}$ + $SiC$
or with $C_{57}Si$ + $C_2$. The first system is  energetically favorable,
but there possibly is high activation energy for that fragmentation process
that does not allow us to see it within our simulation time and statistics.

 A few examples were tried in which some excited (antibonding) 
states of the molecule
were artificially populated, so as to simulate the laser excitation in a 
different way, as 
proposed by other  authors to study non-thermal 
fragmentation \cite{garcia,allen,silvestrelli}.
 However, no qualitatively different results were obtained 
if the excitation energy was small, although after  conversion of the
electronic energy to  phonons, the activated phonons are those in which  the
silicon atoms vibrate more than the carbons, as both the HOMO and the LUMO
have large contributions from the  silicon atoms. Of
course, when all the antibonding states are populated 
a spherically symmetric explosion occurrs.

 Singly substituted $C_{70}$ systems were also studied in some cases, but again no new
fragmentation processes were evident.

\subsection{Coated fullerenes}

 We have also investigated the stable geometries and the dynamical behavior 
 of $C_{60}$ coated with a small number (1, 2, 3, 4, and 6) of $Si$ atoms. In
 the equilibrium configurations our results show that the $Si$ atoms are weakly
 bound to the fullerene surface, each $Si$ atom being located in front of a 
hh $C-C$
   bond and therefore bonded to two $C$ atoms, if the number
 of adsorbed $Si$ atoms is less than three.
  For three adsorbed $Si$ atoms, they are arranged as a
 regular triangle in front of a hexagon.
For four adsorbed $Si$ atoms, one of them has two $C$
 neighbors, of  a hh $C-C$ bond, and the other three have only 
 one $C$ neighbor. For 6 adsorbed $Si$ atoms, two of them have two $C$
 neighbors, two have only one, and the other two have no $C$ neighbors. 
 Table 2 shows the average $Si-Si$ and $Si-C$ bond lengths for the lowest energy
 isomers. In the case of $C_{60}Si_2$ the  $Si$ atoms prefer to be far away from
each other but in systems containing  more than two $Si$ atoms they prefer
 to be close to each other and located in front of a hexagon. These results
 are in good agreement with previous calculations \cite{japoneses}.
 
\begin{table}
\caption{Average $Si-Si$ and $Si-C$ bond lengths for the lowest energy isomers
of fullerenes with adsorbed $Si$ atoms.\\}  
\begin{tabular}{ccc}
& $Si-Si$ ($\AA$)& $Si-C$ ($\AA$) \cr \hline \hline
$C_{60}Si_1$ &  & 2.11 \cr
$C_{60}Si_2$ & 8.01 & 2.11 \cr
$C_{60}Si_3$ & 2.65 & 2.27 \cr
$C_{60}Si_4$ & 2.63 & 2.23 \cr
$C_{60}Si_6$ & 2.62 & 2.26 \cr \hline \hline
\end{tabular}
\end{table}

 In order to study the dynamical behavior  of these $Si$ coated 
 molecules we performed finite temperature molecular dynamics simulations
 with initial
 temperatures of 5000, 4000 and 3000 K. Ten different runs were performed 
 for each system, with 
 different initial maxwellian velocity distributions. Due to the fast 
 interconversion between initial kinetic and potential energies, temperatures 
 of around 2300, 1900  and 1400 K respectively were achieved in less than a 
 picosecond. 
  At these
 temperatures no desorption is observed in systems with one or two $Si$
 atoms, but the adsorbed atoms move  on the fullerene surface. 
 In $C_{60}Si_2$ the two $Si$ atoms move independently and if by chance they 
 collide and form a dimer, they  easily separate  again. 
  As the number of adsorbed $Si$ atoms increases, the desorption probability
 also increases. 
 
\begin{figure}[b]
\epsfxsize=4.7cm\centerline{\epsffile{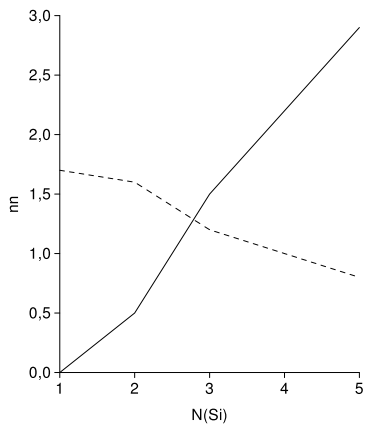}}
\caption{Average number of nearest neighbors ($nn$) of each adsorbed
$Si$ atom in $C_{60}Si_N$, as a function of N(Si), number of adsorbed silicon
atoms}
\label{Fig.3}
\end{figure}

In Table 3 we see 
 that $C_{60}Si_3$ is more stable than $C_{60}Si_4$.  
  The silicon atoms desorb easily from $C_{60}Si_6$  
 as a tridimensional cluster, at even lower temperatures.
 Fig. 3 shows the average number of neighbors of an adsorbed silicon atom
 during the whole simulation time, as a function of the total number of
 adsorbed atoms. If it is less than three they tend to have more carbon than
 silicon neighbors, while
 for  larger groups they form clusters and the number of neighbors  of the same
kind increases.

\begin{figure}
\epsfxsize=6.6cm\centerline{\epsffile{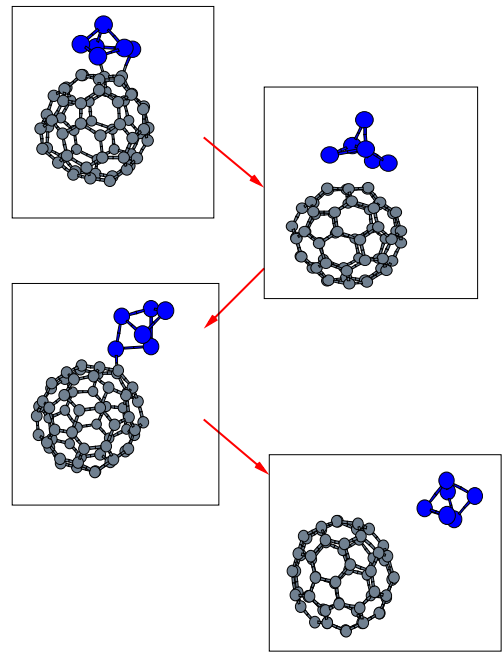}}
\caption{Desorption of the whole adsorbed cluster from $C_{60}Si_6$}
\label{Fig.4}
\end{figure}

 Our simulation results agree with the experimental mass abundance 
 results, that  show a much larger probability of finding molecules with one, 
 two or three adsorbed $Si$ atoms than 
 with four or more  \cite{pellarin3}.
 Another interesting result also shown in Table 3 is that smaller adsorbed
groups
 have  a higher probability  to
 dissociate and then desorb separately 
 while four adsorbed $Si$ atoms or more, tend to desorb
 as a unit (see Fig. 4).
  Fig. 5 shows  $Si_3$ first dissociating into $Si_2$ and $Si$, later 
 $Si_2$ desorbing and the remaining $Si$ adsorbed atom  orbiting on the
 fullerene surface.
 In Fig. 6 another possibility is shown,  two $Si$  atoms  from 
  $C_{60}Si_3$ exchange their positions with two $C$ atoms and as a result they
 are
 incorporated into the fullerene cage. The two $C$ atoms evaporate and therefore the
 final system consists of $C_{58}Si_2$ plus an adsorbed $Si$ atom moving on the
 surface. This  suggests that $Si$ substituted and adsorbed fullerenes may be
 interchangeable cases, with a greater posibility of substituted $Si$ atoms 
 becoming adsorbed ones.  

\begin{figure}[!]
\epsfxsize=6.6cm\centerline{\epsffile{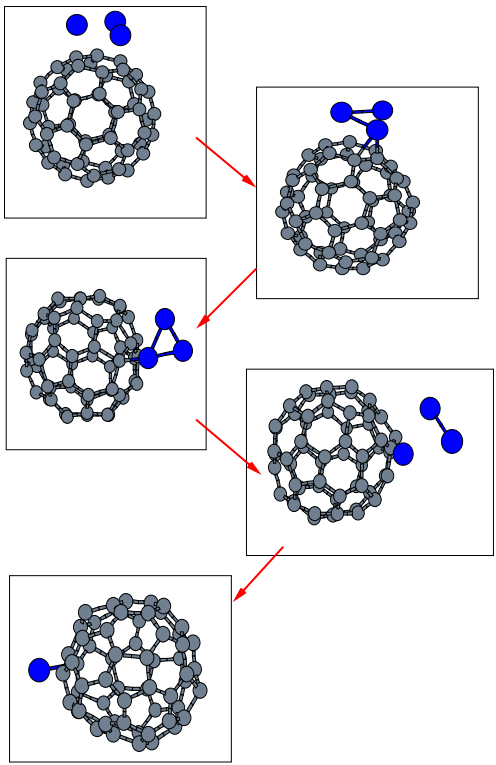}}
\caption{Desorption of $Si_2$ from $C_{60}Si_3$}
\label{Fig.5}
\end{figure}

\section{Discussion and Conclusions}

 Non orthogonal tight-binding calculations are found to reproduce quite 
well the ab-initio static
structural and electronic  results for $Si$ doped fullerenes and  for 
this reason the method seems
 adequate to study the fragmentation of these molecules by 
 molecular dynamics simulations. We have used this method
 in the microcanonical ensemble and simulated the initial excitation by
giving a high initial ionic temperature to the atoms.

 The first fragmentation event in $Si$ substituted fullerenes
shows, as expected, that
 the molecules break in those places where the bond orders are smaller,
 that is, in some $Si-Si$ or $Si-C$ bonds. However, the most frequent first event 
is the ejection
 of a $C_2$ molecule and the second most probable event is that of a $Si$ substituted
 atom going to an adsorbed position. 

For $Si$ coated fullerenes we show that adsorbed $Si$ clusters containing
more than three $Si$ atoms are less stable and tend to desorb as a whole while
smaller clusters
may break into even smaller groups, remain adsorbed or exchange with $C$ atoms
and become substituted fullerenes.
 
 Comparison with experimental results \cite{pellarin2,pellarin3} shows
 several points of agreement, such as the stability of substituted and coated
 fullerenes and the succesive ejection of $C_2$ molecules when excited.
 The difference in abundance between adsorbed clusters with three or more 
 silicon atoms
 is also well reproduced by the simulations.

  The agreements obtained between experiments
and the results of past and also of the present calculations 
suggest that they must contain some real information about the very
beggining of the fragmentation process, on a very local scale where the
total charge and the charge transfers are not the deciding factor. 
 However, the stability of odd numbered molecules and fragments, and the
 fact that the $SiC$ molecule was never found as an ejected fragment 
in the simulations are differences with
  the experiments that remain to be understood.
We plan to continue investigating the reasons
for the discrepancies.

\begin{acknowledgments}

 We thank Dr. Eduardo Hernandez for his help with the
 non-orthogonal tight binding method, 
 providing us with inside information and tables 
of matrix elements. 
 
R.W. acknowledges support from Fundacion Antorchas Grant No.13661-27
\end{acknowledgments}

\begin{figure}[t]
\vskip 1cm
\epsfxsize=6.6cm\centerline{\epsffile{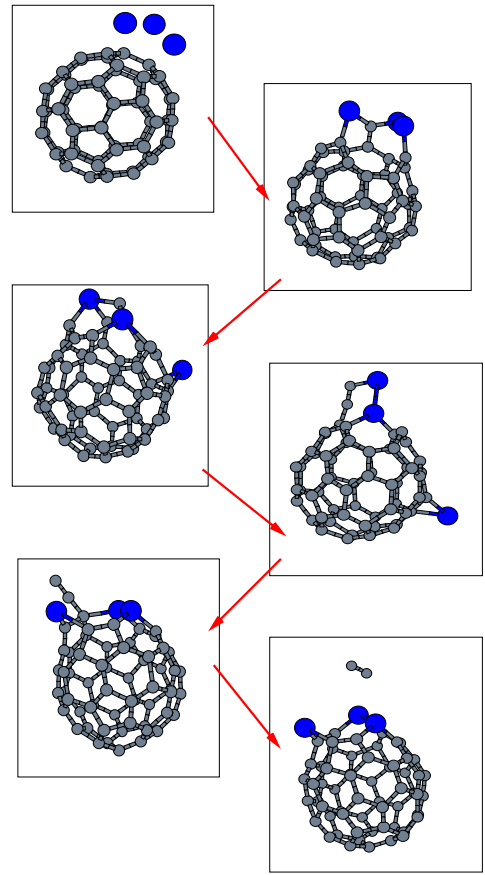}}
\caption{Fragmentation of $C_{60}Si_3$ and exchange of a $Si$ atom with
a $C$ atom} 
\label{Fig.6}
\end{figure}

\begin{widetext}
\begin{center}
\begin{table}
\begin{tabular}{cccc}
System & Initial Temperature   &Frequency &	Event description     \cr
  & &    &  \cr \hline \hline 
& & & \cr
  $C_{60}Si_3$ &5000 K& &      \cr
&  & 0.2 & no desorption within 20 ps     \cr
&  & 0.2 & $Si_3$ ejected between 10 and 20 ps     \cr
&  & 0.1 & $Si_3$ ejected within 10 ps     \cr
&  & 0.3 & $Si_3$ $\rightarrow$ $Si_2$ + $Si$     \cr
&  & 0.2 & exchange of $Si$ with $C$ atom, $ Si$ in the cage    \cr
&  & &      \cr
  $C_{60}Si_4$ & 5000 K & &      \cr
&  & 0.1 & no desorption within 20 ps      \cr
&  & 0.8 & $Si_4$ ejected within 10 ps     \cr
&  & 0.1 & $Si_4$ $\rightarrow$ $Si_3$ + $Si$     \cr
&  & &      \cr
  $C_{60}Si_3$ & 4000 K & &      \cr
&  & 1.0 & no desorption within 20 ps      \cr
&  & &      \cr
  $C_{60}Si_4$ & 4000 K & &      \cr
&  &0.4 & no desorption within 20 ps       \cr
&  &0.4 & $Si_4$ ejected between 10 and 20 ps     \cr
&  &0.2 & $Si_4$ ejected within 10 ps     \cr \hline \hline
\end{tabular}
\caption{Desorption events observed in 20 ps simulation time 
when clusters of three or four $Si$ atoms are adsorbed on the fullerene cage. 
10 simulations were ran in each case}
\end{table} 
\end{center}
\end{widetext}


\begin{references}
\bibitem[*]{fr}
Present Adress:

\bibitem{kroto}
H. W. Kroto, J. R. Heath, S. C. O'Brien, R. F. Curl and R. E. Smalley,
Nature (London) {\bf 318}, 162 (1985)

\bibitem{kimura}
T. Kimura, T. Sugai and H. Shinohara,
Chem. Phys. Lett. {\bf 256}, 269 (1996)

\bibitem{fye}
J. Fye and M. Jarrold,
J.Phys.Chem.A {\bf101}, 1836 (1997)

\bibitem{pellarin1}
C. Ray, M. Pellarin, J. Lerm\'e, J. L. Vialle, M. Broyer, X. Blase,
 P. M\'elinon, P. K\'egh\'elian and A. Perez,
Phys.Rev.Lett {\bf80}, 5365 (1998)

\bibitem{pellarin2}
M. Pellarin, C. Ray, J. Lerm\'e, J. L. Vialle, M. Broyer, X. Blase,
P. K\'egh\'elian, P. M\'elinon and A. Perez,
J.Chem.Phys. {\bf110}, 6927 (1999)

\bibitem{pellarin3}
M. Pellarin, C. Ray, J. Lerm\'e, J. L. Vialle, M. Broyer and P. M\'elinon, 
J.Chem.Phys. {\bf112}, 8436 (2000)

\bibitem{menon}
M. Menon,
J.Chem.Phys. {\bf114}, 7731 (2001)

\bibitem{parrinello}
I. M. L. Billas, C. Massobrio, M. Boero, M. Parrinello, W. Branz, 
F.Tast,
N. Malinowski, M. Heinebrodt and T. P. Martin,
J.Chem.Phys. {\bf111} 6787 (1999)

\bibitem{lu}
J. Lu, Y. Zhou, Y. Luo, Y. Huang, X. Zhang and X. Zhao,
Mol. Phys. {\bf99}, 1203 (2001)

\bibitem{japoneses}
H. Tanaka, S. Osawa, J. Onoe and K. Takeuchi,
J.Phys.Chem.B {\bf103}, 5939 (1999)

\bibitem{nuestro}
Chu-Chun Fu, M. Weissmann, M. Machado and P. Ordejon,
Phys.Rev.B {\bf63} 085411 (2001)

\bibitem{porezag}
D. Porezag, Th. Frauenheim, T. Kohler, G. Seifert, F. Weich and S. Uhlmann,
Phys.Rev.B {\bf 52}, 492 (1995)

\bibitem{frauenheim}
R. Gutierrez, T. Frauenheim, T. Kohler and G. Seifert,
J.Mater.Chem. {\bf6}, 1657 (1996)

\bibitem{nuestroprevio}
Chu-Chun Fu and M. Weissmann,
Phys.Rev.B {\bf60}, 2762 (1999)

\bibitem{eduardo}
S.A. Shevlin, A.J. Fisher and E. Hern\'andez,
Phys.Rev.B {\bf63} 195306 (2001)

\bibitem{scuseria}
C. Xu and G. Scuseria,
Phys.Rev.Lett. {\bf72}, 669 (1994)

\bibitem{tomita}
S. Tomita, J. U. Andersen, C. Gottrup, P. Hvelplund and U. V. Pedersen,
Phys.Rev.Lett. {\bf87}, 073401 (2001)

\bibitem{garcia}
H. O. Jeschke, M. E. Garcia and J. A. Alonso
cond-mat /0104036

\bibitem{allen}
B. Torralva, T. A. Niehaus, M. Elsner, S. Suhai, Th. Frauenheim and R. E. Allen,
Phys.Rev.B {\bf64}, 153105 (2001)

\bibitem{silvestrelli}
A. Gambirasio, M. Bernasconi, G. Benedek and P. L. Silvestrelli,
Phys.Rev.B {\bf62}, 12644 (2000)
\end{references}
\end{document}